\documentclass[11pt,twoside]{article}  % Leave intact
\usepackage{adassconf}
\usepackage{hyperref}

\begin{document}   % Leave intact

%-----------------------------------------------------------------------
%			    Paper ID Code
%-----------------------------------------------------------------------
% Enter the proper paper identification code.  The ID code for your
% paper is the session number associated with your presentation as
% published in the official conference proceedings.  You can           
% find this number locating your abstract in the printed proceedings
% that you received at the meeting or on-line at the conference web
% site; the ID code is the letter/number sequence proceeding the title 
% of your presentation. 
%
% This will not appear in your paper; however, it allows different
% papers in the proceedings to cross-reference each other.  Note that
% you should only have one \paperID, and it should not include a
% trailing period.
%
% EXAMPLE: \paperID{O4-1}
% EXAMPLE: \paperID{P7-7}
%

\paperID{O4a.4, P3.42}

%-----------------------------------------------------------------------
%		            Paper Title 
%-----------------------------------------------------------------------
% Enter the title of the paper.
%
% EXAMPLE: \title{A Breakthrough in Astronomical Software Development}
% 
% If your title is so long as to fill the page header when you print it,
% then please supply a short form as a \titlemark.
%
% EXAMPLE: 
%  \title{Rapid Development for Distributed Computing, with Implications
%         for the Virtual Observatory}
%  \titlemark{Rapid Development for Distributed Computing}
%

\title{Astro-WISE: Chaining to the Universe}
%\titlemark{Astro-WISE}

%-----------------------------------------------------------------------
%		          Authors of Paper
%-----------------------------------------------------------------------
% Enter the authors followed by their affiliations.  The \author and
% \affil commands may appear multiple times as necessary (see example
% below).  List each author by giving the first name or initials first
% followed by the last name.  Authors with the same affiliations
% should grouped together. 
%
% EXAMPLE: \author{Raymond Plante, Doug Roberts, 
%                  R.\ M.\ Crutcher\altaffilmark{1}}
%          \affil{National Center for Supercomputing Applications, 
%                 University of Illinois Urbana-Champaign, Urbana, IL
%                 61801}
%          \author{Tom Troland}
%          \affil{University of Kentucky}
%
%          \altaffiltext{1}{Astronomy Department, UIUC}
%
% In this example, the first three authors, "Plante", "Roberts", and
% "Crutcher" are affiliated with "NCSA".  "Crutcher" has an alternate 
% affiliation with the "Astronomy Department".  The fourth author,
% "Troland", is affiliated with "University of Kentucky"

%\author{Edwin A. Valentijn, John P. McFarland, Jan Snigula\altaffilmark{1}, Kor G. Begeman, Danny R. Boxhoorn, Roeland Rengelink, Ewout Helmich, Philippe Heraudeau, Gijs Verdoes Kleijn, Ronald Vermeij, Willem-Jan Vriend, Michiel J.Tempelaar}
\author{Edwin A. Valentijn, John P. McFarland, Jan Snigula$^\dag$, Kor G. Begeman, Danny R. Boxhoorn, Roeland Rengelink, Ewout Helmich, Philippe Heraudeau, Gijs Verdoes Kleijn, Ronald Vermeij, Willem-Jan Vriend, Michiel J.Tempelaar}
\affil{OmegaCEN, Kapteyn Astronomical Institute, Groningen University, PO Box 800, NL-9700 AV, Groningen, The Netherlands}
\author{Erik Deul, Konrad Kuijken}
\affil{Sterrewacht Leiden, University Leiden, PO Box 9513, NL-2300 RA, Leiden, The Netherlands}
\author{Massimo Capaccioli, Roberto Silvotti}
\affil{INAF, Osservatorio Astronomico di Capodimonte, Salita Moiariello 16, I-80131, Napoli, Italy}
\author{Ralf Bender, Mark Neeser, Roberto Saglia}
\affil{ $^\dag$Universit\"{a}ts-Sternwarte M\"{u}nchen, Scheinerstr. 1, D-81679 M\"{u}nchen, Germany}
\author{Emmanuel Bertin, Yannick Mellier}
\affil{Terapix, Institut d'Astrophysique de Paris, 98bis bd Arago, F-75014, Paris,  France}

%\altaffiltext{1}{Universit\"{a}ts-Sternwarte M\"{u}nchen, Scheinerstr. 1, D-81679 %M\"{u}nchen, Germany}

%-----------------------------------------------------------------------
%			 Contact Information
%-----------------------------------------------------------------------
% This information will not appear in the paper but will be used by
% the editors in case you need to be contacted concerning your
% submission.  Enter your name as the contact along with your email
% address.
% 
% EXAMPLE:  \contact{Dennis Crabtree}
%           \email{crabtree@cfht.hawaii.edu}
%

\contact{Edwin A. Valentijn}
\email{valentyn@astro.rug.nl}

%-----------------------------------------------------------------------
%		      Author Index Specification
%-----------------------------------------------------------------------
% Specify how each author name should appear in the author index.  The 
% \paindex{ } should be used to indicate the primary author, and the
% \aindex for all other co-authors.  You MUST use the following
% syntax: 
%
% SYNTAX:  \aindex{Lastname, F. M.}
% 
% where F is the first initial and M is the second initial (if
% used).  This guarantees that authors that appear in multiple papers
% will appear only once in the author index.  
%
% EXAMPLE: \paindex{Crabtree, D.}
%          \aindex{Manset, N.}        
%          \aindex{Veillet, C.}        
%
% NOTE: this information is also used to build the author list that
% appears in the table of contents.  Authors will be listed in the order
% of the \paindex and \aindex commmands.
%

\paindex{Valentijn, E. A.}
\aindex{Begeman, K. G.}
\aindex{Boxhoorn, D. R.}
\aindex{Deul, E.}
\aindex{Rengelink, R}
\aindex{Helmich, E. M.}
\aindex{Heraudeau, P.}
\aindex{Kuijken, K. H.}
\aindex{McFarland, J. P.}
\aindex{Kleijn, G. V.}
\aindex{Vermeij, R.}
\aindex{Vriend, W.-J.}
\aindex{Tempelaar, M. J.}
\aindex{Capaccioli, M.}
\aindex{Silvotti, R.}
\aindex{Bender, R.}
\aindex{Neeser, M.}
\aindex{Saglia, R.}
\aindex{Bertin, E.}
\aindex{Mellier, Y.}

%-----------------------------------------------------------------------
%		      Author list for page header	
%-----------------------------------------------------------------------
% Please supply a list of author last names for the page header. in
% one of these formats:
%
% EXAMPLES:
% \authormark{Lastname}
% \authormark{Lastname1 \& Lastname2}
% \authormark{Lastname1, Lastname2, ... \& LastnameN}
% \authormark{Lastname et al.}
%
% Use the "et al." form in the case of seven or more authors, or if
% the preferred form is too long to fit in the header.

\authormark{Valentijn et al.}

%-----------------------------------------------------------------------
%			Subject Index keywords
%-----------------------------------------------------------------------
% Enter a comma separated list of up to 6 keywords describing your
% paper.  These will NOT be printed as part of your paper; however,
% they will be used to generate the subject index for the proceedings.
% There is no standard list; however, you can consult the indices
% for past proceedings (http://adass.org/adass/proceedings/).
%
% EXAMPLE:  \keywords{visualization, astronomy: radio, parallel
%                     computing, AIPS++, Galactic Center}
%
% In this example, the author noticed that "radio astronomy" appeared
% in the ADASS VII Index as "astronomy" being the major keyword and
% "radio" as the minor keyword.  The colon is used to introduce another
% level into the index.

\keywords{backward chaining, data processing, data analysis, Virtual Observatory}

%-----------------------------------------------------------------------
%			       Abstract
%-----------------------------------------------------------------------
% Type abstract in the space below.  Consult the User Guide and Latex
% Information file for a list of supported macros (e.g. for typesetting 
% special symbols). Do not leave a blank line between \begin{abstract} 
% and the start of your text.

%\setcounter{footnote}{2}

\begin{abstract}          % Leave intact
% Place the text of your abstract here - NO BLANK LINES
The recent explosion of recorded digital data and its processed derivatives
threatens to overwhelm researchers when analysing their experimental data or
when looking up data items in archives and file systems.  While current
hardware developments allow to acquire, process and store 100s of terabytes of
data at the cost  of a modern sports car, the  software systems to handle these
data are lagging behind.  This problem is very general and is well recognized
by various scientific communities; several large projects have been initiated,
e.g., DATAGRID/EGEE\footnote{\url{http://www.eu-egee.org/}}  federates compute and storage power over the high-energy
physical community, while the international astronomical community is building
an Internet geared Virtual Observatory\footnote{\url{http://www.euro-vo.org/pub/} and Padovani, 2006}  connecting archival data.  These
large projects either focus on a specific distribution aspect or  aim  to
connect many sub-communities and have a relatively long trajectory for  setting
standards and a common layer.  Here, we report ``first light'' of a very
different solution (Valentijn \& Kuijken, 2004) to the problem initiated by a smaller astronomical IT
community.  It provides the abstract ``scientific information layer'' which
integrates distributed scientific analysis with distributed processing and
federated archiving and publishing.  By designing new abstractions and mixing
in old ones, a Science Information System with fully scalable cornerstones has
been achieved, transforming data systems into knowledge systems.  This
break-through is facilitated by the full end-to-end linking of all dependent
data items, which allows full backward chaining from the observer/researcher to
the experiment.  Key is the notion that information is intrinsic in nature and
thus is the data acquired by a scientific experiment.  The new abstraction is
that software systems guide the user to that intrinsic information by forcing
full backward and forward chaining in the data modelling.
\end{abstract}

%-----------------------------------------------------------------------
%			      Main Body
%-----------------------------------------------------------------------
% Place the text for the main body of the paper here.  You should use
% the \section command to label the various sections; use of
% \subsection is optional.  Significant words in section titles should
% be capitalized.  Sections and subsections will be numbered
% automatically. 
%
% EXAMPLE:  \section{Introduction}
%           ...
%           \subsection{Our View of the World}
%           ...
%           \section{A New Approach}
%
% It is recommended that you look at the sample papers, sample1.tex
% and sample2.tex, for examples for formatting references, footnotes,
% figures, equations, html links, lists, and other special features.  

\section{Introduction}

The classical paradigm to handle data streams of large physical experiments,
such as CERN-HEP, astronomical telescopes, or scientific satellites, is
characterized by different layers in the project that apply certain algorithms
to the stream of data and subsequently deliver the results to the next layer,
following a so-called TIER architecture.  This architecture can be
characterized as data-driven and ``feed-forward''.  The construction of the
different layers is often grown historically and relates to implicit or
explicit project management decisions facing sociology, geography and
interfacing of working groups with different expertise.  Projects operated
under this paradigm work with releases of datasets, which are obtained with a
certain version of the code meeting certain quality control standards.
Operators have a task to push the input data through the stream, often by means
of a ``pipeline'' irrespective of whether the derived data items are actually
used by the end users.  Examples of successful forward workflow projects are
numerous (e.g., Valentijn \& Kuijken, 2004; Costa et al., 2003; Hiriart et al., 2003; Jung et al., 2003; Mehringer \& Plante 2003; and Pierfederici et al., 2003).  When funding for the operation centres dries
up, a final release of processed data is made, and the project comes to an end.

\begin{figure}
\epsscale{.80}
\plotone{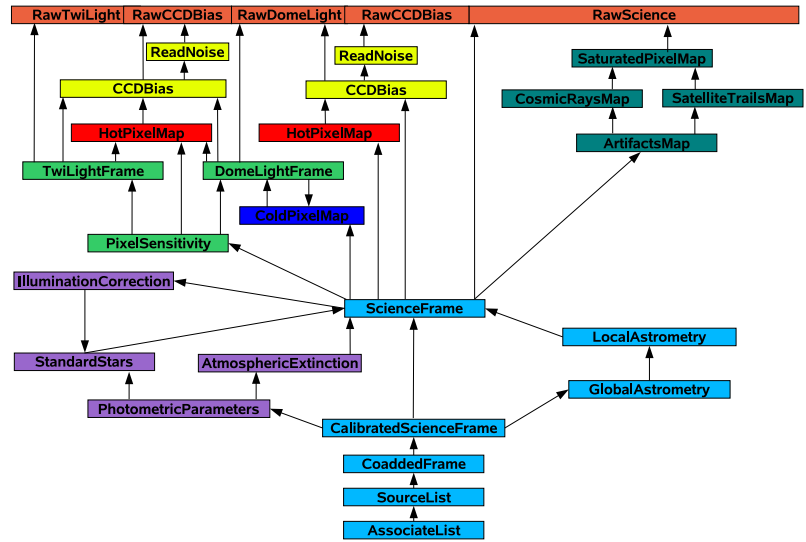}
\caption{A {\it target diagram}: slightly simplified view of the dependencies
of ``targets'' to the ocean of raw observational data of astronomical wide
field imaging. Arrows indicate the backward chaining to the raw data.}
% The diagram contains two circular
%dependencies (cold pixel map, Illumination correction) which can be supported.
%The optional, tunable ``depth'' used during processing of a target refers to the
%number levels of up-stream located dependencies which are checked for
%``up-to-dateness''.  Note that this depth can be much larger than suggested by
%this simplified view.}
\label{fig1}
\end{figure}

\section{The Problem}

This ``classical paradigm'' has the advantage of publishing complete sets of
data with well described methods, qualifications, and calibrations.  However,
it has bad scalability when exposed to modern huge data streams: re-processing
all the raw data and storing the results for new releases when new
computational methods, calibration strategies, insights or improved code
becomes available is very difficult to impossible.  And, moreover, why should
operators re-process all data as long as it is even not known  whether there
are customers for the individual products?  Though the classical approach has
the obvious advantage of facilitating established releases, it cannot
facilitate the demands of the end-user/researcher, who actually might want to
evaluate specific questions, and specific results obtained with specific
methods from (a subset)  in the flood of physical observational data (including
even holiday digital pictures or scanned documents of national libraries) with.
Key to addressing the merits of a system for a researcher is to evaluate the
question of what analysis on uncertainties the researcher facilitates in
that system before (s)he submits his/her 4 sigma result to a scientific
journal.

Such demands combined with the increased data rates from instruments and
world-wide communities ask for a different approach.  A first interesting step
is set by the Space Telescope Science Institute, operating the Hubble Space
Telescope, where users can ask for a certain data product which will then run
the Calibration pipeline on-demand using the best available calibrations (Swade et al., 2001), a
case of forward chaining.  But what to do when a user discovers a very faint
object and wants to inspect the robustness of the result by checking the
dependencies on uncertainties of calibration parameters and applied methods?
The user wants to interact with the data, derive his own result following his
insights and methods, and preferably this knowledge is fed back into the system
at the disposal of other researchers who optionally want to take advantage of
the progress of  insight.  In the 1980s, solutions were explored in expert
systems.  Nowadays, with distributed communities working on enormous data
floods, new artificial intelligence-like, distributed information systems are
required to facilitate the scientific endeavour.

\section{The Solution}

A new generation of wide field astronomical imaging cameras includes the
MEGACAM at the CFHT and OmegaCAM (Kuijken et al., 2004) at the VLT Survey Telescope (VST). OmegaCAM, with a
one square degree field of view and a pixel size of 0.2 arc second, will
deliver 256 mega-pixel images of the sky every few minutes, for 300 nights per
year for many years to come.  These experiments will produce 100's of terabyte
per year of raw and processed data, e.g., to trace the effects of dark matter
and dark energy.  These requirements inspired an international consortium,
Astro-WISE, to design and implement a completely different approach to connect
the end-user to the experiment.  Facing the problems sketched above we  build
an information system in which the role of the user (observer) is central.  The
system is built to handle queries by the user for his/her desired result, which
we call a ``target''.  A ``target'' could be an astrometrically and
photometrically calibrated image, or the results for a set of calibration
parameters, or a list of parameter values describing an astronomical object, or
whatever ``target'' that is facilitated by the system.  In the abstractions used
for the software the ``target'' processing resembles that of the UNIX make
metaphor, but in addition the dynamic aspects resembles  to the ``goal'' in
artificial intelligence systems.

\section{How it Works}

Following a query by the user, the system checks whether the ``target''  has been processed before,
\begin{itemize}
\item if not, it will be processed ``on-the-fly'', recursively following all
its dependencies in an object model, similar to the Unix make metaphor.  See
the ``target diagram'' in Fig. \ref{fig1}.
\item if already existent the system will check all its dependencies up to the
raw data taken at the telescope (Fig. \ref{fig1})
\begin{itemize}
\item if all its dependencies are ``up-to-date'' the target object is returned.
\item if all or some dependencies are not ``up-to-date'' they will be
re-computed on-the-fly, again according to the `make' metaphor, but with
optional, tunable ``depth''.
\end{itemize}
\end{itemize}

All data production in the system follows this schema, whether it concerns an
individual user asking for a single special result (target), a calibration
scientist determining the instrument behaviour over long periods or a
production scientist deriving results for whole nights of observing.  They all
add knowledge  to the system.  Note that, in the end, it does not matter very
much whether or not a target requested by the user has been processed before,
it only affects the  performance of the system.  This way, the workflow in the
ocean of data space is fully driven by the users quest.  (see
\url{http://process.astro-wise.org/})

This behaviour is achieved by linking (referencing) all dependent data items in
the information system.  This is achieved by carefully specifying the
observation procedure both for science and calibration observations and
subsequently designing a data model, chaining the processing targets to the raw
data.  Next, the data model is converted into an object model, which in turn is
ported in a commercial database (using object oriented ``user-defined types'' in
a relational database).  The database thus has full awareness of all
dependencies.  When compute scripts are run at a computer, all Class
instantiations are automatically made persistent in the database forming a
dynamic archive of all targets and added knowledge, while new and nearly
``sacred'' raw data is ingested as long as the acquisition continues.  Table \ref{table1} summarizes some key differences between the Astro-WISE system and the classical approach.

\begin{table}
\caption{Characteristics of the Astro-WISE Information System}
\begin{center}
\begin{tabular}[ht]{|l|l|}
\hline\hline
``Classical'' Paradigm & Astro-WISE Paradigm \\
\hline
forward chaining      & backward chaining       \\
``tier'' architecture & ``target'' architecture \\
driven by raw data    & driven by user query    \\
process in pipeline   & process on-the-fly      \\
operators push data   & users pull data         \\
results in releases   & information system      \\
static archive        & dynamic archive         \\
raw data is obsolete  & raw data is sacred      \\
\hline
\end{tabular}
\end{center}
\label{table1}
\end{table}

\begin{figure}
\epsscale{.80}
\plotone{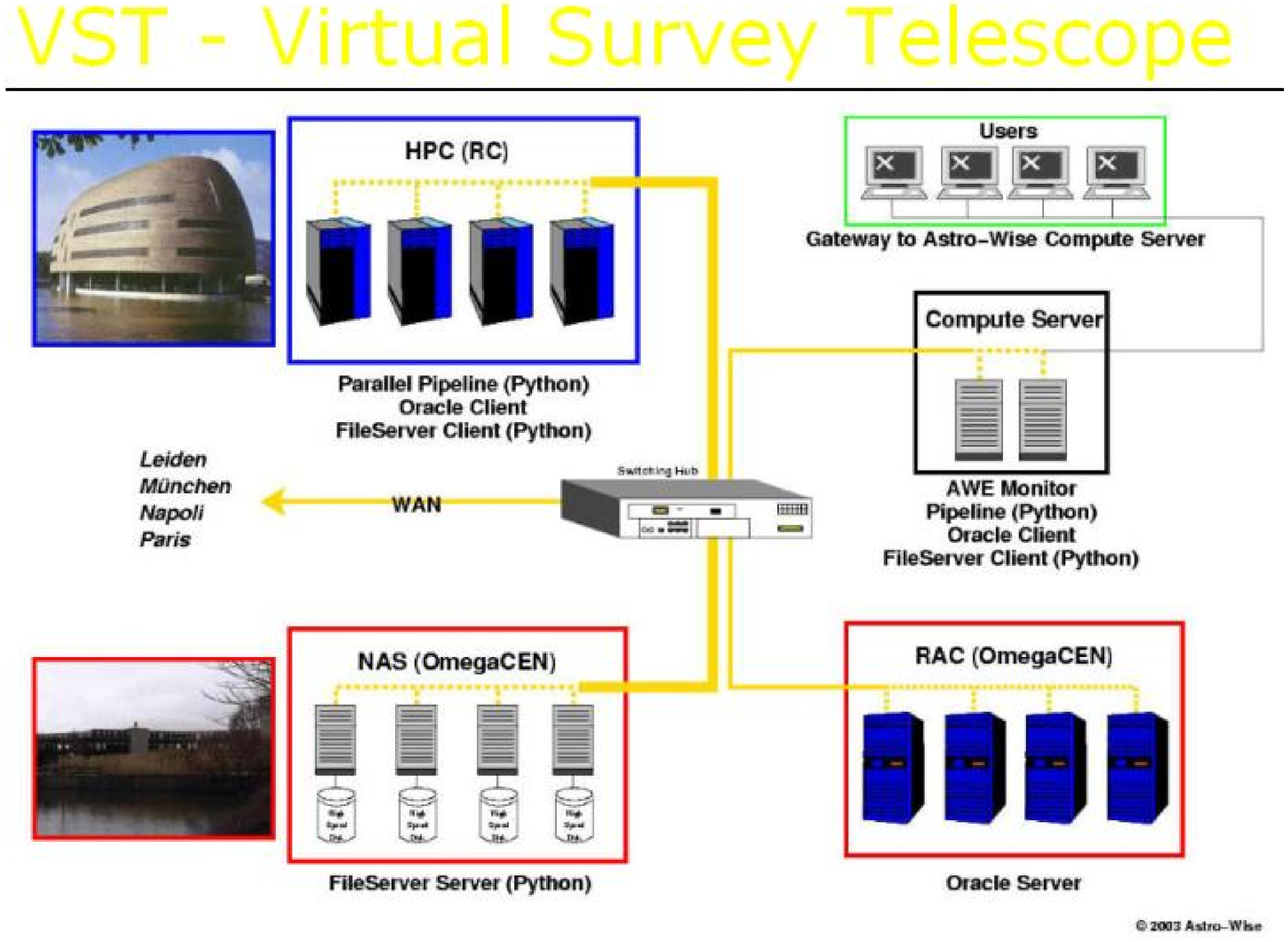}
\caption{From virtual computer observing to real computer observing. The figure
sketches the four fully scalable cornerstones for computer observing in a peer
to peer network (with international federations).}%:
%\begin{enumerate}
%\item users,
%\item Linux farm for embarrassingly parallel processing,
%\item database cluster (Oracle Real Application Cluster), and
%\item storage area network.
%\end{enumerate}
%Roles of clients and servers are indicated.  All input/output is controlled by
%the database.  At point ii. any Linux machine with an IP address can be
%connected for processing and is initiated in 10 sec (similar to SETI\@home).
%At point iii. all meta-data and astronomical source data is within the database
%and is scalable up to 100's Tb using Oracle 10g Partitioning.  At point iv. all
%image data items are on a SAN controlled by federated fileserver.}
\label{fig2}
\end{figure}

Obviously, this can only be achieved by integrating all computing and storage
hardware into a single information system.  Figure \ref{fig2} shows the
peer-to-peer network employed for Astro-WISE.  The network and its hardware can
be viewed as an extension of the telescope and measurement apparatus hardware,
together with the abstract software, forming a unity.  The user, by requesting
a target, triggers the whole chain from target to raw observational data of the
measurement apparatus, a real virtual observatory.

All top level software in the system is written in the Python scripting
language (see \url{http://www.python.org/}) which in turn calls C libraries.
Following the LISP language, Python achieves a very high level of integration.
It facilitates another crucial unification in the system: apart from the GUIs,
all usage of the system is done from a Python prompt via a Python binding to
the database.  Novice users, advanced users, and scientific
programmers/developers alike use the same environment, allowing a continuous
transition between all levels.  All Python open source facilities, e.g.,
visualisation libraries (MatPlotLib), numerical libraries (NumPy), etc., are at
direct disposal of the user and are  connected to the output of user queries to
the information system.  Users apply scripts provided by a federated CVS
distribution of the standard Astro-WISE code base, but can also modify scripts,
fine-tuning them or adding methods as long as they do not violate the object
model.  This way, both a simple standard version and user tuned versions of
processing targets can be obtained simultaneously (different versions of the
code are tagged and can be traced).

The database is partitioned in instruments and projects, facilitating
individual read and write privileges for persons, groups and projects.  Users
set a context relating to their project and thus see only their partition in
the ocean of data.  The smallest project, the individual or anonymous Web based
user doing his/her ``afternoon'' experiment is facilitated in a ``MyDB''
context, which can be upgraded to another, appropriate context/project at the
end of the session.  Project leaders have the responsibility for the project
quality control.  The code base supports this by built-in ``verify'',
``compare'' and ``inspect'' methods for each Class, where ``verify'' involves
an internal code check, ``compare'' involves a database query to different
instantiations of the same Class and ``inspect'' is a visual inspection by the
user. Eventually, all quality control is maintained by a time-stamping
mechanism on each Class instantiation, and a GUI is build for calibration
scientists to supervise and alter the timestamps, the ultimate point where
human insights add knowledge to the system.

In fact, an important design criterion was to allow a complete federation of
the system, facilitating different research groups spread over Europe to share
scientific projects.  Currently, the system connects National data centres in
the Netherlands (Groningen), Italy (Naples), France (Paris), and Germany
(Munich), which in turn re-distribute the system to satellites (e.g., Bonn and
Heidelberg).  Similar hardware is installed at the National centres: the
code-base is federated, the database is replicated, and the file-servers prevent
duplication of massive storage of image data.  Network traffic is minimised by
allowing for terabyte-sized local cache for frequently used objects.  All GUIs
are web based.  These include the database viewer
(\url{http://dbview.astro-wise.org/}) and the GUI to evaluate and operate target
processing at any selected Linux farm of the federation.  User (client) altered
Python scripts are sent to the processors in the federation as so-called Python
``pickles'' (persistent Python objects in a string representation).

The system has been fully implemented and currently operates the raw data of
various astronomical wide field imaging cameras at ESO's 2.2m telescope in
Chile (300.000 CCD read-outs), the ING telescope(Canary Islands), the Subaru
telescope (Japan) and OmegaCAM for ESO's VLT survey telescope (currently test
data only).  During the implementation, while taking care of object model
purity, we experienced an avalanche of benefits for which relatively little
extra effort had to be made:

\begin{itemize}
\item The full linking of object dependencies allows full backwards chaining as
employed by the artificial intelligence (Shachter \& Heckerman, 1987 and Thompson \& Thompson, 1985) and history tracking.  Literally
every bit of information that went into the target can be retrieved: the system
provides a ``tell me everything tool''.  In fact, this is an a priori
ontological implementation, in contrast to the fashionable a posteriori,
semantic web search engines.

\item Publishing on Internet/EURO-VO (the European Virtual Observatory) is done
by raising a flag.  It is up to project managers to do this for classical
paradigm static results or for Astro-WISE paradigm dynamic results.

\item The control by the database over the  parallel processing (e.g.,
SETI@home) on any number of nodes.The enabling of a international compute GRID. 

\item Enabling global astrometric and photometric solutions with increased
accuracy, redirecting global database knowledge back into the system.  

\item The built-in workflow directly guides the user and no workflow systems are required.
\end{itemize}

Next to facilitating statistical studies, the system is optimized for
``needle-in-the-haystack'' kind of searches, finding extremely rare
astronomical objects, such as moving, solar system objects, or variable objects
like ultra-compact binary systems of white dwarf stars or distant supernovae.
Next to all image Meta-data, the astronomical source parameters are stored in
an Oracle database and Oracle partitioning is used to quickly address up to 100
Tbyte data volumes.  Fast astronomical object associations (a posteriori
associations) are made in ``many-to-many'' mode using both native database
indexing and positional HTM\footnote{\url{http://www.sdss.jhu.edu/htm/}} indexing.
The linking (joins and references) are maintained at an extreme level.  For
example, in the KIDs 1500-5000 degree survey nearly 1 Tbyte of linked data
items is anticipated.

The avalanche of side products and its rapid implementation are thanks to
several unifications  which are  achieved by merging various pointer
mechanisms, such as those provided by object oriented programming
(inheritance), class/attribute persistency,  database internal links (joins,
references), and namespace handling by the Python scripting language.  In the
context of physical information theory, all these links can be viewed as kinds
of memory addresses and facilitate forms of information transfer, gearing the
user to intrinsic information, and facilitating the handling of the dispersion
caused by the measurement apparatus, camera, and the off-line computing
hardware.  Particularly, the option to inspect partial derivatives of dependent
parameter, by re-deriving results, allows the user to inspect the information
content in the data (Frieden, 1998).

\section{Conclusion \& Acknowledgements}

Our key-concept, involving novel approaches to maintain data associations in
federations of integrated pipelines and archives, can be applied to arbitrary
forms of digitized observational data, ranging from DNA sequences to numerical
simulations and national libraries (e.g., centuries of Dutch governmental
handwritten records being scanned and entered into the system will be processed
with pattern recognition techniques).  After all, this is all part of our
Universe that we observe, and with the coming decade, will be copied into our
hardware at multi-petabyte scales to be interpreted by a  next generation of
scientists.

The project received  funding from the European Union  through the EC Action
{\it Enhancing Access to Research Infrastructures}, a FP5 RTD programme, the
Netherlands Research School for Astronomy NOVA and NWO.  Correspondence and requests for materials should be
addressed to E.A. Valentijn, Coordinator, Astro-WISE {\it valentyn@astro.rug.nl}.

\noindent
Supplementary Information can be found on \url{http://www.astro-wise.org/}

\end{document}